\documentclass[fleqn,usenatbib]{mflds}

\usepackage{newtxtext,newtxmath}

\usepackage[T1]{fontenc}
\newcommand{\bl}[1]{\mbox{\boldmath$ #1 $}}

\usepackage{graphicx}
\usepackage{amsmath}
\usepackage[version=4]{mhchem}
\usepackage{bm}

\title[The Effect of Luminosity Outbursts]{The Effect of Luminosity Outbursts on the Abundance of Pebbles and Their Ice Mantles in Protoplanetary Disks}
\author[A. Topchieva et al.]{
A. Topchieva,$^{1}$\thanks{E-mail: ATopchieva@inasan.ru}
T. Molyarova,$^{2}$
and E. Vorobyov$^{1,2}$
\\
$^{1}$Institute of Astronomy, Russian Academy of Sciences, 48 Pyatnitskaya St., Moscow, 119017, Russia\\
$^{2}$Research Institute of Physics, Southern Federal University, Rostov-on-Don, Russia
}
\date{}
\begin{document}

\maketitle

\begin{abstract}
{
Dust growth is one of the key processes leading to planet formation in protoplanetary disks. Centimeter-sized dust grains---pebbles---are necessary for planetesimal formation via the streaming instability, they play an important role in forming protoplanetary cores and giant planets, as well as enriching their atmospheres with chemical elements. This work investigates the effect of luminosity outbursts on the abundance of pebbles and their ice mantles in protoplanetary disks. We perform global simulations of formation and evolution of a self-gravitating viscous protoplanetary disk using the 2D hydrodynamic thin-disk FEOSAD code, which self-consistently reproduces luminosity outbursts. The model includes thermal balance, dust evolution and its interaction with gas, development of magnetorotational instability, adsorption and desorption of four volatile compounds (H$_2$O, CO$_2$, CH$_4$ and CO), and the feedback of ice mantles on dust fragmentation properties. We show that luminosity outbursts have a stronger effect on the positions of CO$_2$, CH$_4$ and CO snowlines compared to the water snowline. This is because the H$_2$O snowline falls within the viscous heating dominated region during early disk evolution stages, while snowlines of other molecules are located in regions dominated by stellar irradiation heating and are thus more sensitive to temperature changes during outbursts. Nevertheless, luminosity outbursts reduce the total amount of pebbles in the disk by half due to destruction of dust aggregates into monomers following the loss of water ice that binds the aggregates together. Pebble recovery occurs over several thousand years after the outburst ends due to collisional coagulation, with recovery timescales significantly exceeding water freeze-out times. Ice mantle desorption occurs in a complex non-axisymmetric 2D region of the disk, associated with spiral substructure formation during early evolution of gravitationally unstable disks.}
\vspace{0.3cm}
\end{abstract}

\maketitle

\section{Introduction}

Dust growth to centimeter sizes is one of the main processes leading to planet formation in protoplanetary disks. Such large dust grains, also called pebbles, facilitate the formation of protoplanetary cores and giant planets through pebble accretion \citep{2015Icar..258..418M,2017AREPS..45..359J}. Pebbles are also necessary for planetesimal formation, for example in the streaming instability scenario \citep{2005ApJ...620..459Y,2015A&A...579A..43C}. Thus, pebble properties must be studied to understand planet formation conditions.

Dust grain sizes in disks depend on many processes, with collisional evolution being one of the most important~\citep{2014prpl.conf..339T}. In this process, dust coagulation is counteracted by fragmentation, parameterized by the fragmentation velocity $v_{\rm frag}$, which is the maximum collision velocity between dust grains that leads to coagulation rather than fragmentation. Typical $v_{\rm frag}$ values are $1-10$\,m~s$^{-1}$ \citep{2005PhRvE..71b1304W,2008ARA&A..46...21B,Steinpilz2019ApJ...874...60S,Pillich2021A&A...652A.106P}.

\citet{2020A&A...637A...5E} and \citet{Vorobyov2023A&A} showed that pebble quantity and spatial distribution in protoplanetary disks significantly depend on fragmentation velocity. The fragmentation velocity in turn depends on dust grain properties, particularly the presence of ice mantles that weaken fragmentation~\citep{2009ApJ...702.1490W,2015ApJ...798...34G,2016ApJ...821...82O}. Ice mantles on dust grain surfaces are thought to facilitate the formation of pebbles.

During disk evolution, ice mantles protecting dust grains from fragmentation can evaporate. Grains lose their mantles when they drift radially \citep{1977MNRAS.180...57W} into warmer inner disk regions inside the water snowline \citep{2011ApJ...735..131S}. Additionally, many young stellar objects experience FU~Ori-type luminosity outbursts that can heat the disk and evaporate ice mantles beyond typical snowline locations \citep{2016Natur.535..258C,2017A&A...604A..15R}.
Ice mantle destruction during young star outbursts leads to potentially observable changes in protoplanetary disk dust properties~\citep{Vorobyov2022,2023MNRAS.521.5826H}.

\citet{2024MNRAS.530.2731T} previously showed that pebbles form within the first tens of thousands of years after protoplanetary disk formation and are covered by ice composed mainly of H$_2$O and CO$_2$. Luminosity outbursts self-consistently occurring in FEOSAD modeling affect ice composition and should alter dust properties in the disk, including pebbles.
\citet{Vorobyov2022} showed that luminosity outbursts heat protoplanetary disks, affecting radial positions of H$_2$O snowlines, shifting them by up to several tens of au. Under the adopted ``many seeds'' model \citep{2017A&A...602A..21S}, this process leads to dust aggregate matrix disruption into monomers following the loss of water ice that binds aggregates together.
This paper describes the effect of luminosity outbursts on the mass of pebbles and composition of their ice mantles based on \citet{2024MNRAS.530.2731T} simulations. Unlike \citep{Vorobyov2022} where idealized luminosity outbursts were manually imposed, we use self-consistent modeling of accretion and luminosity outbursts resulting from magnetorotational instability development in inner disk regions. Moreover, this work focuses on outburst effects on pebble spatial distribution and evolution rather than the entire dust size spectrum.

The paper is structured as follows. Section~\ref{sec:results111} ``Model Description'' presents the numerical model and parameters, emphasizing new features compared to \cite{Vorobyov2018A&A...614A..98V}: gas and dust dynamics (Section~\ref{sec:dust_growth}), viscosity parameterization (Section~\ref{sec:visc_model}), pebble definition (Section~\ref{sec:criteria}). Section~\ref{sec:resalttt} presents main simulation results describing luminosity outburst effects on pebbles and their ice mantles (Section~\ref{sec:fluxresalttt}). Section~\ref{sec:theend} summarizes conclusions about the composition of ice on pebbles ice and its dynamics in protoplanetary disks during luminosity outbursts.

\section{Model Description}
\label{sec:results111}

This work uses numerical simulations of protostellar/protoplanetary disk formation and evolution with the FEOSAD
(Formation and Evolution Of Stars And Disks) model presented in \citet{Vorobyov2018A&A...614A..98V}, with subsequent modifications in \citet{Molyarova2021ApJ}. Detailed model description can also be found in \citet{2024MNRAS.530.2731T}. FEOSAD is a 2D hydrodynamic code in the thin-disk approximation. Simulations begin with gravitational collapse of a prestellar core and end at the T Tauri stage when protostar and protoplanetary disk form. The simulation covers spatial scales from $0.6$\,au to several thousand au with radially increasing logarithmic spatial resolution in inner disk regions, enabling study of various disk structures like spiral arms, rings and self-gravitating clumps.

One advantage of such 2D modeling is that unlike viscous evolution equations of the Pringle type \citep[see][]{Pringle1981}, it uses full hydrodynamic equations while requiring significantly less computational resources compared to full 3D approaches. This thin-disk model serves as an excellent tool for studying long-term protoplanetary disk evolution. The FEOSAD code includes disk self-gravity, turbulent viscosity parameterized via the $\alpha$-approach \cite{Shakura1973A&A}, radiative cooling and heating by stellar and background radiation, viscous heating, dust evolution, self-consistent dust drift relative to gas (not simplified as in viscous evolution equations), dust back reaction on gas, and dynamics and phase transitions of key volatiles.

\subsection{Gas and Dust Dynamics}
\label{sec:dust_growth}

Hydrodynamic equations describing mass, momentum and internal energy conservation for the gas component are:

\begin{equation}
\frac{{\partial \Sigma_{\rm g} }}{{\partial t}} + \nabla \cdot
\left( \Sigma_{\rm g} \bl{v} \right) =0,
\label{eq:1}
\end{equation}

\begin{equation}
\frac{\partial}{\partial t} \left( \Sigma_{\rm g} \bl{v} \right) + \nabla \cdot \left( \Sigma_{\rm
g} \bl{v} \otimes \bl{v} \right) = - \nabla {\cal P} + \Sigma_{\rm g} \, \bl{g}
+ \nabla \cdot \mathbf{\Pi} - \Sigma_{\rm d,gr} \bl{f},
\label{eq:2}
\end{equation}

\begin{equation}
\frac{\partial e}{\partial t} +\nabla \cdot \left( e \bl{v} \right) = -{\cal P}
(\nabla \cdot \bl{v}) -\Lambda +\Gamma + \nabla \bl{v}:{\bl \Pi},
\label{energ}
\end{equation}

where $\Sigma_{\rm g}$ is gas surface density, $e$ is internal energy per unit surface area, $\cal P$ is vertically integrated gas pressure from the ideal gas equation ${\cal P} = (\gamma -1)e$ with $\gamma = 7/5$,
$\bl{f}$ is gas-dust friction force, $\bl{v} = v_{r}\hat{r}+v_{\phi}\hat{\phi}$ is gas velocity in the disk plane, $\nabla = \hat{\bl r} \partial / \partial r + \hat{\bl \phi}r^{-1} \partial / \partial \phi$ is the planar gradient. Gravitational acceleration in the disk plane $\bl g = g_{r}\hat{\bl r} + g_{\phi}\hat{\bl \phi}$ includes central protostar gravity and gas/dust self-gravity calculated by solving the Poisson integral~\citep{Vorobyov2010ApJ...719.1896V}.
Expressions for radiative cooling $\Lambda$ and radiative heating $\Gamma$ (the latter including stellar and background radiation) can be found in \cite{Vorobyov2018A&A}. Background radiation temperature is set to 15\,K. In FEOSAD, turbulent viscosity is included via the viscous stress tensor $\bl{\Pi}$. Viscosity parameterization is described in more detail in Section~\ref{sec:visc_model}.

Gas and dust co-evolve and interact via gravitational and friction forces, the latter including dust back reaction on gas following the analytical method by \cite{Stoyanovskaya2018ARep}. Initially all dust in the cloud is in submicron form but grows as the protoplanetary disk forms and evolves. Small submicron dust is assumed to be coupled to the gas, while grown dust can dynamically decouple from gas. This paper describes the dust growth scheme presented in \cite{Molyarova2021ApJ}. In FEOSAD, small and grown dust populations are represented by their surface densities $\Sigma_{\rm d,sm}$ and $\Sigma_{\rm d,gr}$, respectively. Each dust population has a power-law size distribution $N(a) = Ca^{-p}$ with normalization constant $C$ and fixed exponent $p = 3.5$. For small dust, minimum size $a_{\rm min}= 5 \times 10^{-7}$\,cm, maximum size $a_\ast = 10^{-4}$\,cm. For grown dust, $a_\ast$
is the minimum size and $a_{\rm max}$ is the maximum size that can change via dust coagulation and fragmentation, as well as advective transport through the disk.

Dynamics of small and grown dust grains is described by the following continuity and momentum equations (small dust is strictly dynamically coupled to gas):

\begin{equation}
\label{contDsmall}
\frac{{\partial \Sigma_{\rm d,sm} }}{{\partial t}} + \nabla \cdot
\left( \Sigma_{\rm d,sm} \bl{v} \right) = - S(a_{\rm max}) + \nabla \cdot \left( D \Sigma_{\rm g} \nabla\left(\frac{\Sigma_{\rm d,sm}}{\Sigma_{\rm g}} \right)\right),
\end{equation}
\begin{equation}
\label{contDlarge}
\frac{{\partial \Sigma_{\rm d,gr} }}{{\partial t}} + \nabla \cdot
\left( \Sigma_{\rm d,gr} \bl{u} \right) = S(a_{\rm max}) + \nabla \cdot \left( D \Sigma_{\rm g} \nabla\left(\frac{\Sigma_{\rm d,gr}}{\Sigma_{\rm g}} \right)\right),
\end{equation}
\begin{equation}
\label{momDlarge}
\frac{\partial}{\partial t} \left( \Sigma_{\rm d,gr} \bl{u} \right) + \nabla \cdot \left( \Sigma_{\
d,gr} \bl{u} \otimes \bl{u} \right) = \Sigma_{\rm d,gr} \, \bl{g}
+ \Sigma_{\rm d,gr} \bl{f} + S(a_{\rm max}) \bl{v},
\end{equation}

where $\bl u$ is grown dust velocity.
$S(a_{\rm max})$ represents the small-to-grown dust conversion rate per unit disk surface area as dust grows, i.e., as maximum dust size $a_{\rm max}$ increases (in~g~s$^{-1}$~cm$^{-2}$) \citep[see details in][equation (14)]{2020A&A...644A..74V}. The diffusion coefficient is calculated from kinematic viscosity $D= \nu / Sc$, with Schmidt number $Sc=1$. Friction force $\bl{f}$ calculation is described in \cite{2020ARep...64..107S,Vorobyov2023A&A}.

In the dust growth model, maximum grown dust size $a_{\rm max}$ is calculated at each timestep for each cell as the solution to
\begin{equation}
{\frac{\partial a_{\rm max}}{\partial t}} + (u \cdot \nabla ) a_{\rm max} = \cal{D},
\label{dustA}
\end{equation}
where dust growth rate $\cal{D}$ describes $a_{\rm max}$ change due to coagulation, and the second left-hand term describes $a_{\rm max}$ change from dust advection between cells. We calculate $\cal{D}$ as
\begin{equation}
\cal{D}=\frac{\rho_{\rm d} {\it v}_{\rm rel}}{\rho_{\rm s}},
\label{GrowthRateD}
\end{equation}
where $\rho_{\rm d}$ is dust volume density in the midplane, $\rho_{\rm s}=3$\,g~cm$^{-3}$ is refractory dust core density, and $v_{\rm rel}$ is dust grain collision velocity. The main collision sources are assumed to be Brownian motion and turbulence determined by the Shakura-Sunyaev $\alpha$ parameter. This approach is described in more detail in \citet{Vorobyov2018A&A...614A..98V}.

The $a_{\rm max}$ value is limited by the fragmentation barrier~\citep{2012A&A...539A.148B}; maximum dust size cannot exceed $a_{\rm frag}$ determined by dust properties. If $a_{\rm max}$ exceeds $a_{\rm frag}$ at any point, $\cal{D}$ is set to zero and $a_{\rm max}$ is set to $a_{\rm afrag}$. The maximum fragmentation-limited dust size is
\begin{equation}
a_{\rm frag}=\frac{2\Sigma_{\rm g}v_{\rm frag}^2}{3\pi\rho_{\rm s}\alpha c_{\rm s}^2},
\label{eq:afrag}
\end{equation}
where $v_{\rm frag}$ is the fragmentation threshold depending on ice mantle presence (see Section~\ref{sec:chem_model}). Note the dust drift barrier is self-consistently included via coupled gas-dust dynamics with mutual friction.

\subsection{Turbulent Viscosity Parameterization}
\label{sec:visc_model}

The model implements a variable Shakura-Sunyaev $\alpha$-parameter used to calculate turbulent viscosity. Viscosity, significantly contributing to mass and angular momentum transport in protoplanetary disks, is primarily considered as caused by turbulence generated by magnetorotational instability \citep{Balbus1991ApJ, Turner2014prpl}. In FEOSAD, turbulent viscosity is included via the viscous stress tensor $\bl \Pi$ (see equations ~\eqref{eq:2},~\eqref{energ}), with kinematic viscosity $\nu = \alpha c_{\rm s} H$, where $c_{\rm s}$ is sound speed and $H$ is the vertical scale height of the gas calculated assuming local hydrostatic equilibrium. To mimic accretion through a layered disk, we consider an effective adaptive $\alpha_{\rm eff}$ parameter \citep[Eq. (12) from][]{Kadam2022MNRAS}. The $\alpha_{\rm eff}$ parameter is calculated as a weighted average for combined active and dead magnetorotational instability layers:
\begin{equation}
\alpha_{\rm eff}=\frac{\Sigma_{\rm MRI}\alpha_{\rm MRI}+\Sigma_{\rm dz}\alpha_{\rm dz}}{\Sigma_{\rm MRI}+\Sigma_{\rm dz}}.
\end{equation}
The active layer sufficiently ionized by cosmic rays for magnetorotational instability development has surface density $\Sigma_{\rm MRI}=200$\,g~cm$^{-2}$ (100\,g~cm$^{-2}$ from each disk surface), while the dead zone has $\Sigma_{\rm dz}=\Sigma_{\rm g}-\Sigma_{\rm MRI}$. The viscosity parameter in the active layer is $\alpha_{\rm MRI}=10^{-3}$, while in the dead zone $\alpha_{\rm dz}=10^{-5}$ for temperatures below $1300$\,K and $\alpha_{\rm dz}=10^{-1}$ otherwise, which corresponds to magnetorotational instability development from thermal ionization of alkali metals in inner disk regions.
This $\alpha_{\rm eff}$ parameterization helps approximately account for turbulence suppression in the so-called dead zone and corresponding reduction in mass transport, as well as accretion and luminosity outbursts from episodic magnetorotational instability development in inner hot disk regions. More detailed descriptions can be found in \cite{Bae2014ApJ, Kadam2019ApJ, Kadam2022MNRAS}.

Luminosity outbursts can be triggered by various mechanisms causing temporary increases in stellar accretion rates. A key mechanism is magnetorotational instability~\citep[MRI,][]{Balbus1991ApJ,1998RvMP...70....1B,Turner2014prpl} development in inner disk regions. Corresponding $\alpha$ growth leads to accretion outbursts, or MRI outbursts~\citep{2010ApJ...713.1134Z,2020ApJ...895...41K,2020A&A...644A..74V}. This mechanism is self-consistently implemented in the presented model, leading to episodic growth of accretion luminosity. Each considered model experiences several dozen such outbursts during simulation, with amplitudes and durations similar to FUor outbursts \citep{2014prpl.conf..387A,2018ApJ...861..145C}. These outbursts affect disk thermal structure, which in turn influences dust and pebble properties.

\subsection{Pebble Definition and Identification}
\label{sec:criteria}
The Stokes number in protoplanetary disks typically describes dust dynamics. It is defined as:
\begin{equation}
St = \frac{{ \Omega_{\rm k} \rho_{\rm s} a_{\rm max}}}{{ \rho_{\rm g} c_{\rm s}}}
\end{equation}
where $\Omega_{\rm k}$ is Keplerian angular velocity, $\rho_{\rm g} = \Sigma_{\rm g} / \sqrt{\rm 2 \pi} H$ is midplane gas volume density. Grains with $St\sim$1 are most strongly affected by gas and rapidly drift radially toward pressure maxima. Millimeter-to-centimeter sized dust grains called pebbles play a crucial role in planet formation scenarios involving pebble accretion \citep{Ormel2010, Lambrechts2012, Ida2016A&A, Lambrechts2017}. Pebbles have relatively high $St>0.01$ and move relative to gas, unlike submicron grains that are coupled to gas and move together with it.

This work uses the pebble definition from \citet{Vorobyov2023A&A}, with modifications also described in \citet{2024MNRAS.530.2731T}. FEOSAD considers two dust populations, small ($a<a_{\rm *}=1$\,$\mu$m) and grown ($a>a_{\rm *}$) dust. We define pebbles as grown dust with Stokes number and grain radius exceeding threshold values $St>St_{\rm 0}$ and $a_{\rm peb,0}$. In disk regions where both conditions are met, pebbles have a size distribution that is part of the grown dust size distribution $N(a)= C a^{-p}$, ranging from $a_{\rm peb,min}$ to $a_{\rm max}$. The minimum pebble size $a_{\rm peb,min}$ is:
\begin{equation}
a_{\rm peb,min}=
\begin{cases}
a_{\rm St_{\rm 0}}, & \text{ if } a_{\rm St_{\rm 0}} > a_{\rm peb,0}, \\
a_{\rm peb,0}, & \text{ if } a_{\rm St_{\rm 0}} \leq a_{\rm peb,0}.
\end{cases}
\end{equation}
Here $a_{\rm St_{\rm 0}}$ is the grain size with Stokes number $St_{\rm 0}$. In regions containing pebbles, it is defined as
\begin{equation}
a_{\rm St_{\rm 0}}=a_{\rm max}\frac{St_{\rm 0}}{St}.
\end{equation}

We adopt $St_{\rm 0}=0.01$ and $a_{\rm peb,0}=0.05$ based on studies in \cite{Lambrechts2012, Lenz2019ApJ}. The Stokes number identifies large grains moving independently from gas, but it also depends on local gas density. Our simulation includes an envelope surrounding the disk where $St$ reaches high values for micron-sized grains due to low gas density. The minimum pebble size condition $a_{\rm peb,0}$ excludes these small grains from consideration.

Pebble surface density $\Sigma_{\rm peb}$ in each computational cell is calculated assuming its size distribution continues the grown dust size distribution $N(a)= C a^{-p}$, leading to:
\begin{equation}
\Sigma_{\rm peb} = \frac{{ \Sigma_{\rm d, gr} \left({\sqrt{a_{\rm max}} - {\sqrt{a_{\rm peb, min}}}} \right) }}{{\sqrt{a_{\rm max}} - {\sqrt{a_{\rm *}}}}}
\end{equation}

Pebbles defined this way are part of grown dust within our two-population dust model constraints. Their dynamics is indistinguishable from overall grown dust dynamics described by equations (\ref{contDlarge}) and (\ref{momDlarge}). To distinguish these populations, we consider grown dust surface density minus pebble surface density unless stated otherwise. Ice surface densities on pebbles are calculated assuming ice mass fractions on pebbles equal those on grown dust.

Importantly, dust destruction at sudden $v_{\rm frag}$ drops is instantaneous in our modeling. This stems from the implementation of dust growth rate $\cal{D}$ and $v_{\rm frag}$ described in Section \ref{sec:dust_growth}. When mantles evaporate, the fragmentation barrier $a_{\rm frag}$ changes abruptly, and all dust exceeding the new $a_{\rm frag}$ is recycled mainly into small dust plus some larger grains determined by the new $a_{\rm frag}$. This instantaneous destruction scenario is justified if grains are aggregates of small grains stuck together by ice mantles. Then mantle disappearance should instantly disrupt grains \citep[see e.g.][]{2017A&A...605L...2S}. An alternative approach explicitly includes collisional destruction (erosion), as in \citet{2022ApJ...935...35S}. In that case destruction is not instantaneous but occurs on coagulation timescales of order $10^{3}$\,years \citep{Vorobyov2022}, potentially longer than outburst durations. That scenario assumes ice mantles completely cover silicate grain cores.
Both scenarios were considered in \citet{2024MNRAS.527.9668H}, and the comparison to V883~Ori spectral indices favors prolonged dust destruction in that object. However, in self-gravitating protoplanetary disks with prominent spiral structure, complex snowline geometry should lead to mantle-stuck aggregates as shown in \citet{Molyarova2021ApJ}. This motivates more detailed consideration of different grain disruption scenarios. Within FEOSAD this could be implemented via a bidisperse dust model including erosion as described in \citet{2020MNRAS.499.5578A}, planned as future work extending this study.

\subsection{Chemical Model Description}
\label{sec:chem_model}

The numerical simulation considers four volatile molecules: H$_2$O, CO$_2$, CH$_4$ and CO, which can exist in gas, on small dust and on grown dust. When we identify pebbles as part of grown dust in post-processing, we can separately consider ices on pebbles. Distribution of volatiles in the disk and between phases changes via three main processes. First, they advect with gas and dust. Second, during collisional dust evolution ice mass redistributes between dust fractions similarly to refractory core mass redistribution: if a certain mass fraction of small dust converts to grown dust, the same fraction of each ice type on small dust converts to ice on grown dust. We also consider phase transitions: adsorption onto dust grain surface, thermal desorption and photodesorption. The volatile evolution model is described in more detail in \cite{Molyarova2021ApJ}.

The model assumes that initially, small dust grains are covered by ice mantles, grown dust has not yet formed, and gas-phase volatiles are absent. Initial relative molecular abundances are based on ice observations in low-mass protostellar cores \citet{Oberg2011a} with ratios H$_2$O : CO$_2$ : CO : CH$_4$ $=100:29:29:5$. Initial ice mass fraction relative to refractory material is $\approx$8.5\%. Protoplanetary disks can have different abundances of volatiles. Since volatile mass is not included in dust dynamics, initial composition choice will not qualitatively affect the results. It may influence ice ratios but not snowline positions or dust properties.

Adopted initial abundances characterize prestellar core compositions representing the stage immediately preceding protostar and protoplanetary disk formation, they are often used for astrochemical modeling \citep{2016A&A...595A..83E,2018A&A...613A..14E}. Carbon elemental abundance in this chemical composition is reduced relative to solar, implying the presence of solid carbonaceous dust and refractory organics in the disk not included in current modeling. Accounting for these components and their possible transition to gas phase requires considering chemical processes beyond desorption and adsorption, which is beyond the scope of this work, but could potentially affect gas composition in the inner disk \citep{2010ApJ...710L..21L,2019ApJ...870..129W}.

To estimate the effect of ice mantles on dust evolution we use fragmentation threshold $v_{\rm frag}$ as a parameter depending on ice mantle presence on grown dust. At collision velocities below $v_{\rm frag}$ grown grains stick together, above it they fragment. We assume that ice mantles generally increase $v_{\rm frag}$ regardless of composition \citep[][see]{Molyarova2021ApJ}, based on laboratory studies \cite{2009ApJ...702.1490W, 2015ApJ...798...34G}. Ice-covered grains have $v_{\rm frag}=5$\,m~s$^{-1}$, bare grains have $v_{\rm frag}=0.5$\,m~s$^{-1}$. Grown dust in a disk region is considered ice-covered if there is enough ice on grown dust to cover all present grown grains by at least one ice monolayer (thickness order of molecular size).
Monolayer thickness is estimated using characteristic water molecule size $3 \times 10^{-8}$\,cm. After calculating ice surface densities at each timestep, $v_{\rm frag}$ values are updated.

\subsection{Model Parameters}
\label{sec:m1m2_model}

We consider two models with different initial cloud masses $\rm M_{\rm core} =$ 0.66\,$M_{\odot}$ and 1.0\,$M_{\odot}$. Model M1 serves as baseline, while M2 investigates how model mass affects pebble property conclusions. Disk-to-star mass ratios are relatively high and similar (0.55 and 0.60), making both disks gravitationally unstable, but total system mass also affects gas and dust evolution. Central star mass in the model is self-consistently determined by accretion flow from the disk. Over 0.5\,Myr the disk retains significant mass, maintaining gravitational instability and thus inward mass transport needed to activate magnetorotational instability \citep{Bae2014ApJ}.

At 0.5\,Myr the disk remains massive, which is reproduced in other numerical protoplanetary disk formation models \cite[e.g.][]{2020ApJ...891..154D}. Observational determination of disk masses mainly relies on indirect methods using CO and grown dust emission, and potentially underestimates masses by 1-2 orders of magnitude due to optical depth effects and model uncertainties \citep{2014MNRAS.444..887D, 2014A&A...572A..96M, 2017ApJ...849..130M}. More direct methods like gas kinematics yield values comparable to our simulations \citep[e.g.][]{2023MNRAS.518.4481L}.
Main model parameters are listed in Table~\ref{tab:model}. The simulations used a $N_r \times N_{\phi} = 390 \times 256$ grid. Disk formation occurs at 53\,kyr in model M1 and 79\,kyr in model M2.

\begin{table*}
\caption{Model parameters. $M_{\rm core}$ is initial core mass, $\beta$ is the ratio of rotational to gravitational energy, $T_{\rm init}$ is initial gas temperature equal to background radiation temperature $T_{\rm bg}$, $\Omega_0$ is characteristic core rotation angular velocity, $\Sigma_0$ is central gas surface density, $r_0$ is central plateau radius, $R_{\rm out}$ is outer boundary. $M_{\rm star}$ and $M_{\rm disk}$ are central star and disk masses at the end of the simulation (500\,kyr). Our model assumes that $\approx$10\% of accreted mass is ejected in jets and outflows. Little mass remains in the envelope by the end of the simulation.}
\label{tab:parameters}
\begin{tabular}{lccccccccc}
\hline\noalign{\smallskip}
Model & $M_{\rm core}$ & $\beta$ & $T_{\rm init}$ & $\Omega_0$ & $\Sigma_0$ & $r_0$ & $R_{\rm out}$ & $M_{\rm star}$ (0.5 Myr) & $M_{\rm disk}$ (0.5 Myr) \\
& ($M_{\odot}$) & (\%) & (K) & (km s$^{-1}$ pc$^{-1}$) & (g cm$^{-2}$) & (au) & (au) & ($M_{\odot}$) & ($M_{\odot}$) \\
\hline\noalign{\smallskip}
M1 & 0.66 & 0.28 & 15 & 2.26 & $1.73\times 10^{-1}$ & 1029 & 6116 & 0.40 & 0.22 \\
M2 & 1.00 & 0.28 & 15 & 1.51 & $1.15\times 10^{-1}$ & 1543 & 9170 & 0.58 & 0.35 \\
\noalign{\smallskip}\hline
\label{tab:model}
\end{tabular}
\end{table*}

\section{Results and Conclusions}
\label{sec:resalttt}

\begin{figure*}
\includegraphics[width =\columnwidth]{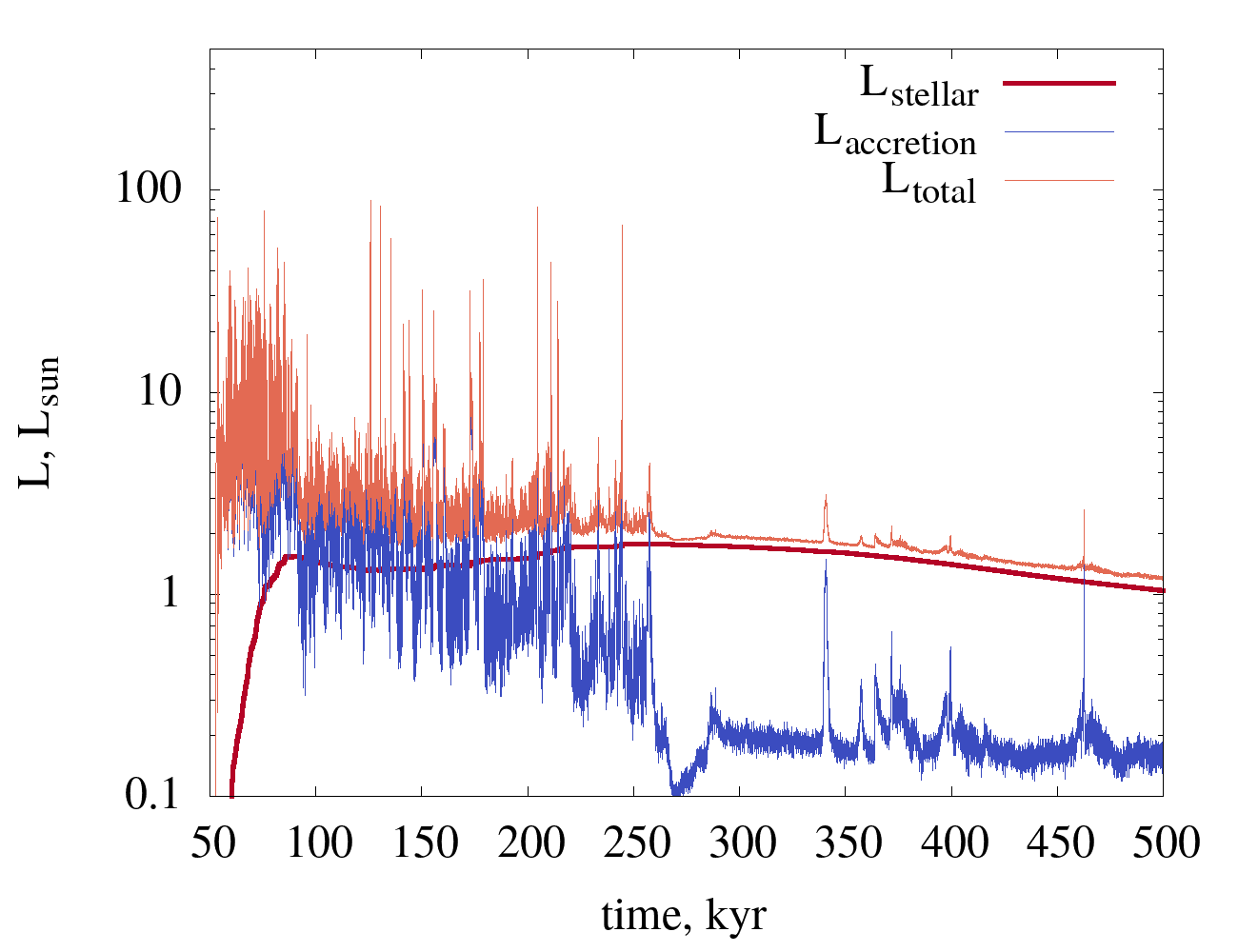}
\includegraphics[width =\columnwidth]{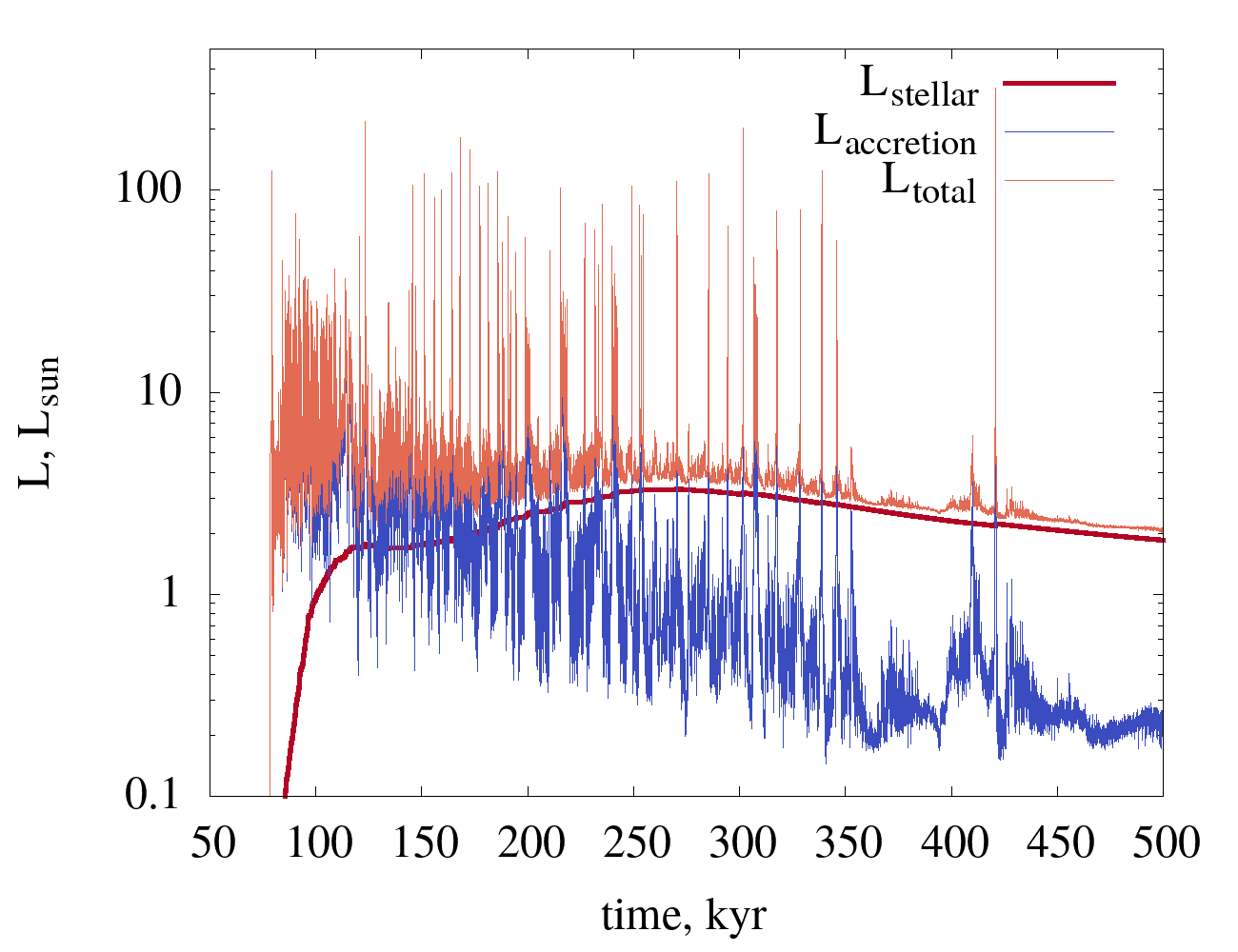}
\caption{Stellar luminosity and accretion luminosity in models M1 (left) and M2 (right).}
\label{ris:accretion pattern}
\end{figure*}

Beyond $\geq 10$\,au, disk temperature is largely determined by stellar radiative heating depending on accretion and photospheric luminosity. Stellar accretion and photospheric luminosity evolution over disk lifetime in both models is shown in Fig.~\ref{ris:accretion pattern} left panel. Accretion outbursts with magnitudes of tens to hundreds $L_{\odot}$ occur regularly during the first few hundred thousand years of disk evolution. The characteristics of the outbursts differ between models.
In M1 they nearly cease after 250~kyr. In M2 they continue until 350~kyr, with an isolated bright outburst at $\sim425$~kyr. Longer activity in the more massive model results from prolonged fueling of inner disk regions where instability and accretion outbursts develop.
Moreover, M2 outbursts have higher magnitudes $\sim100-200$~$L_{\odot}$ versus $\sim30-80$~$L_{\odot}$ in M1.

\subsection{Luminosity Outburst Effects on Pebbles and Ices}
\label{sec:fluxresalttt}

\begin{figure*} 
\includegraphics[width =1\columnwidth]{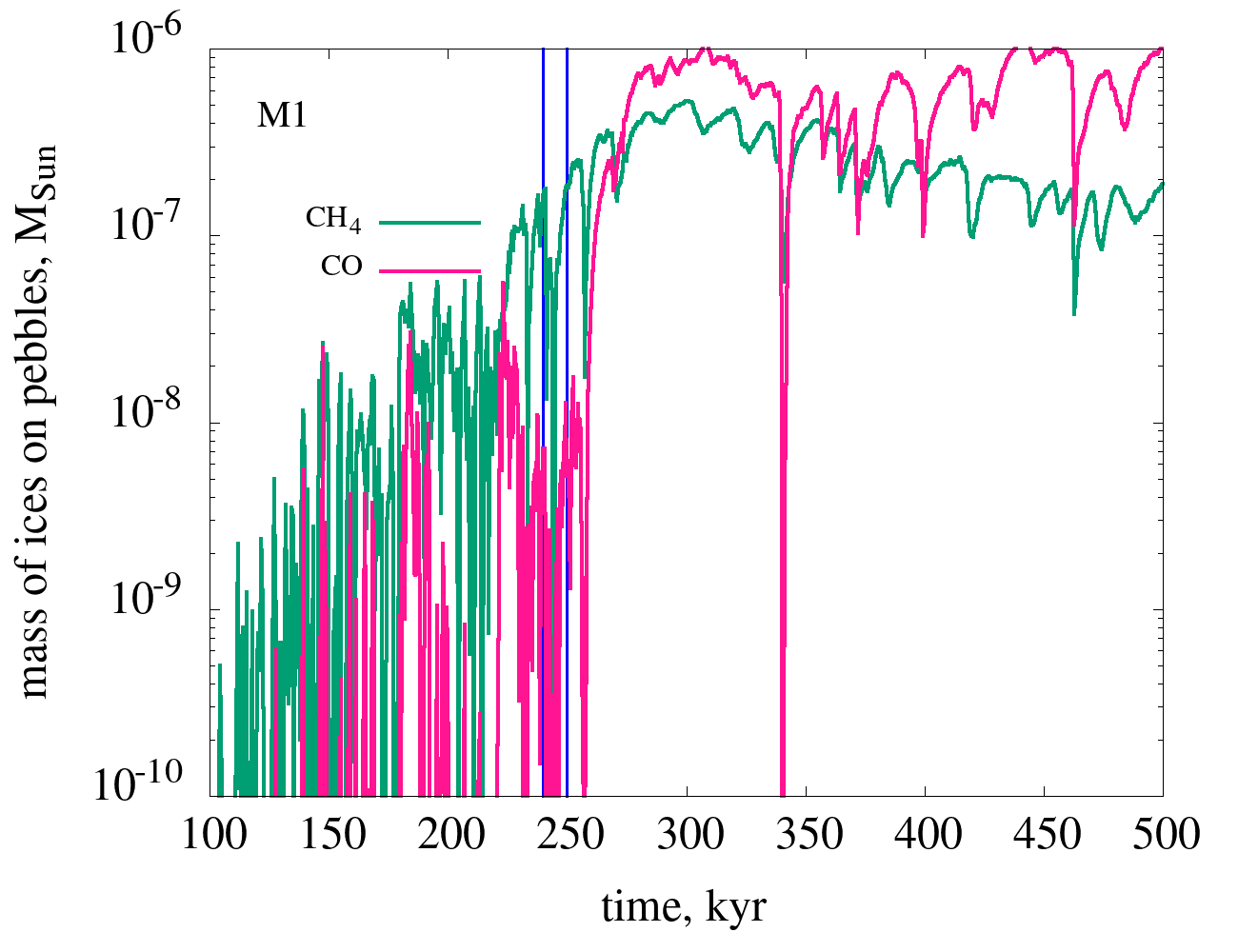}
\includegraphics[width =1\columnwidth]{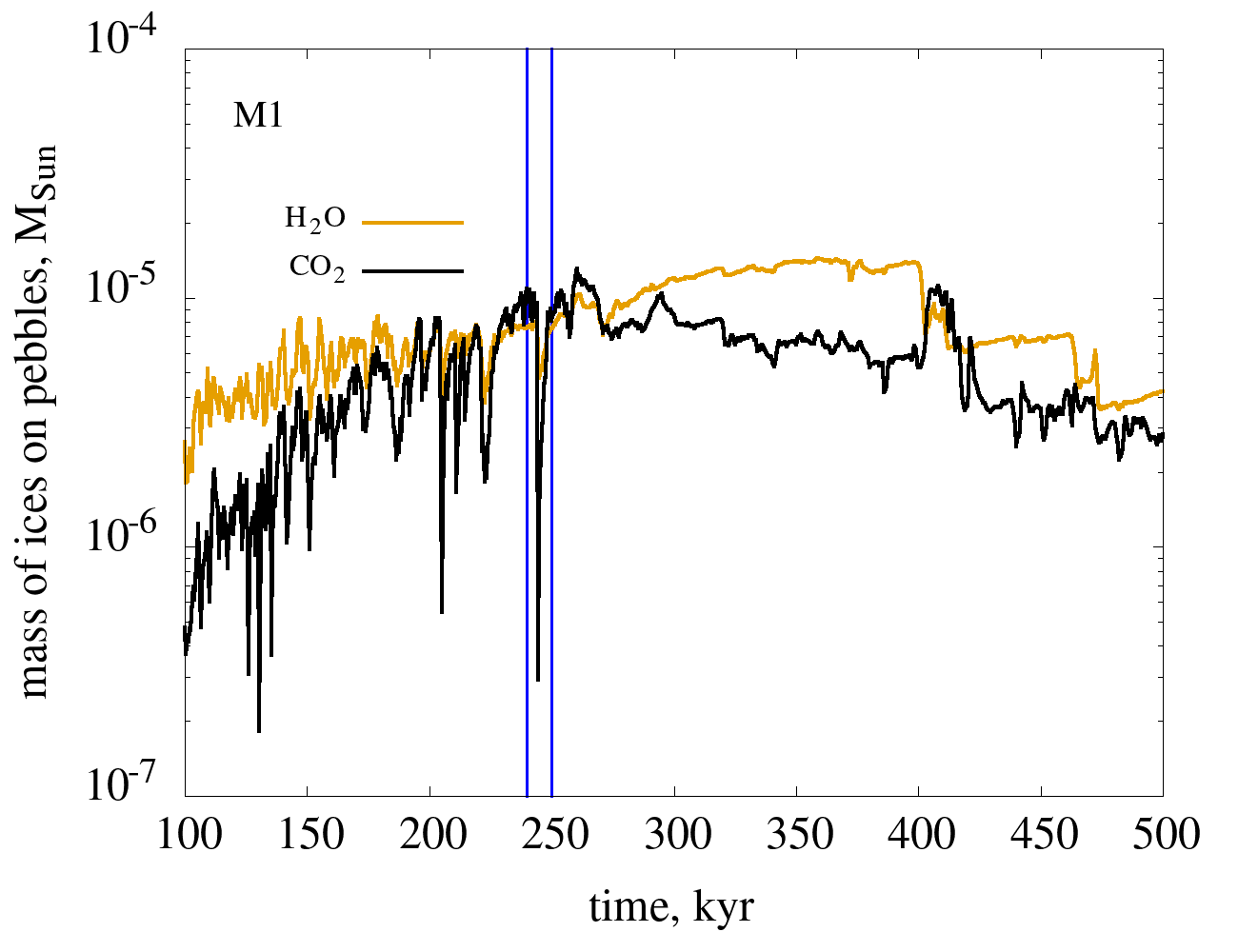} \\
\includegraphics[width =1\columnwidth]{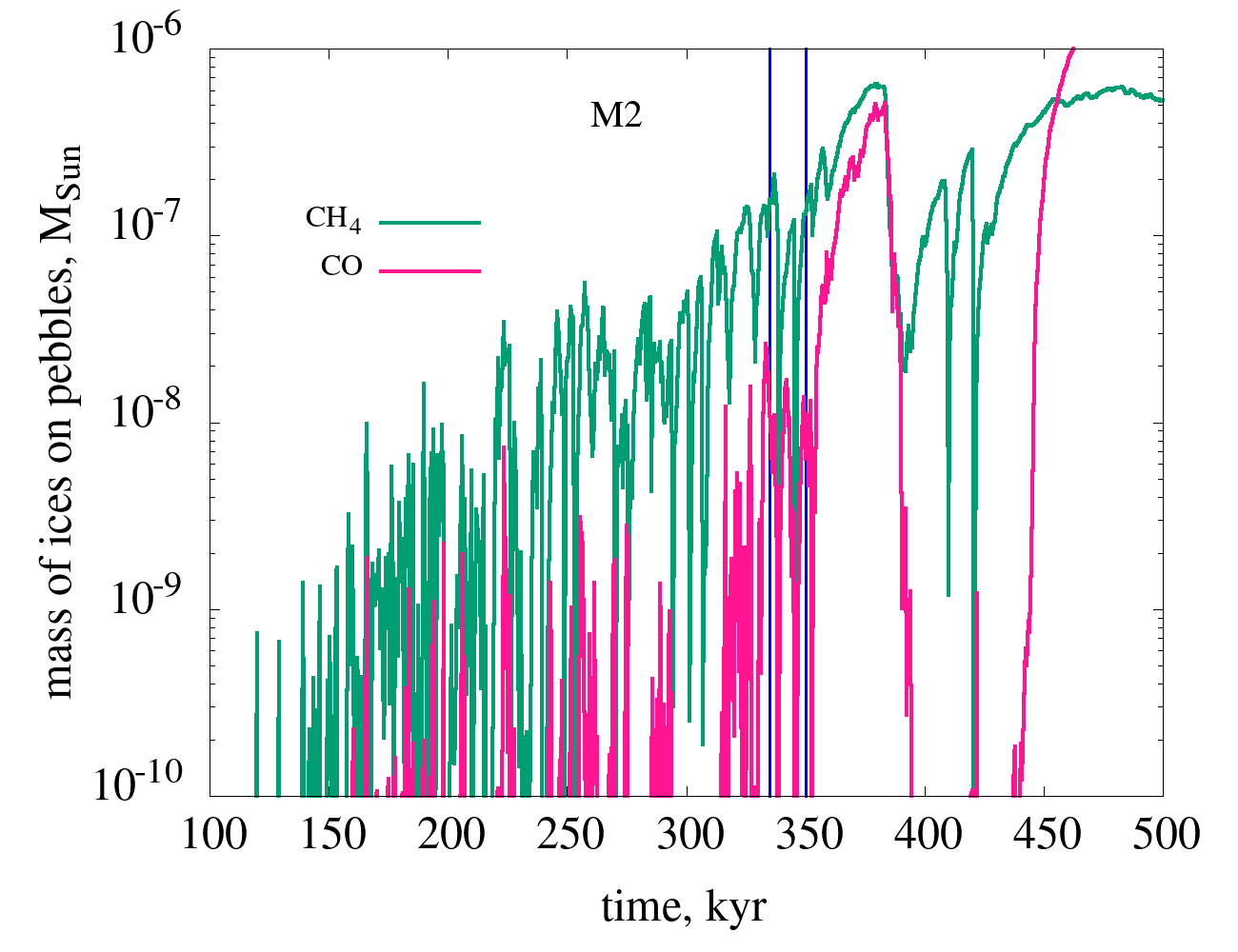}
\includegraphics[width =1\columnwidth]{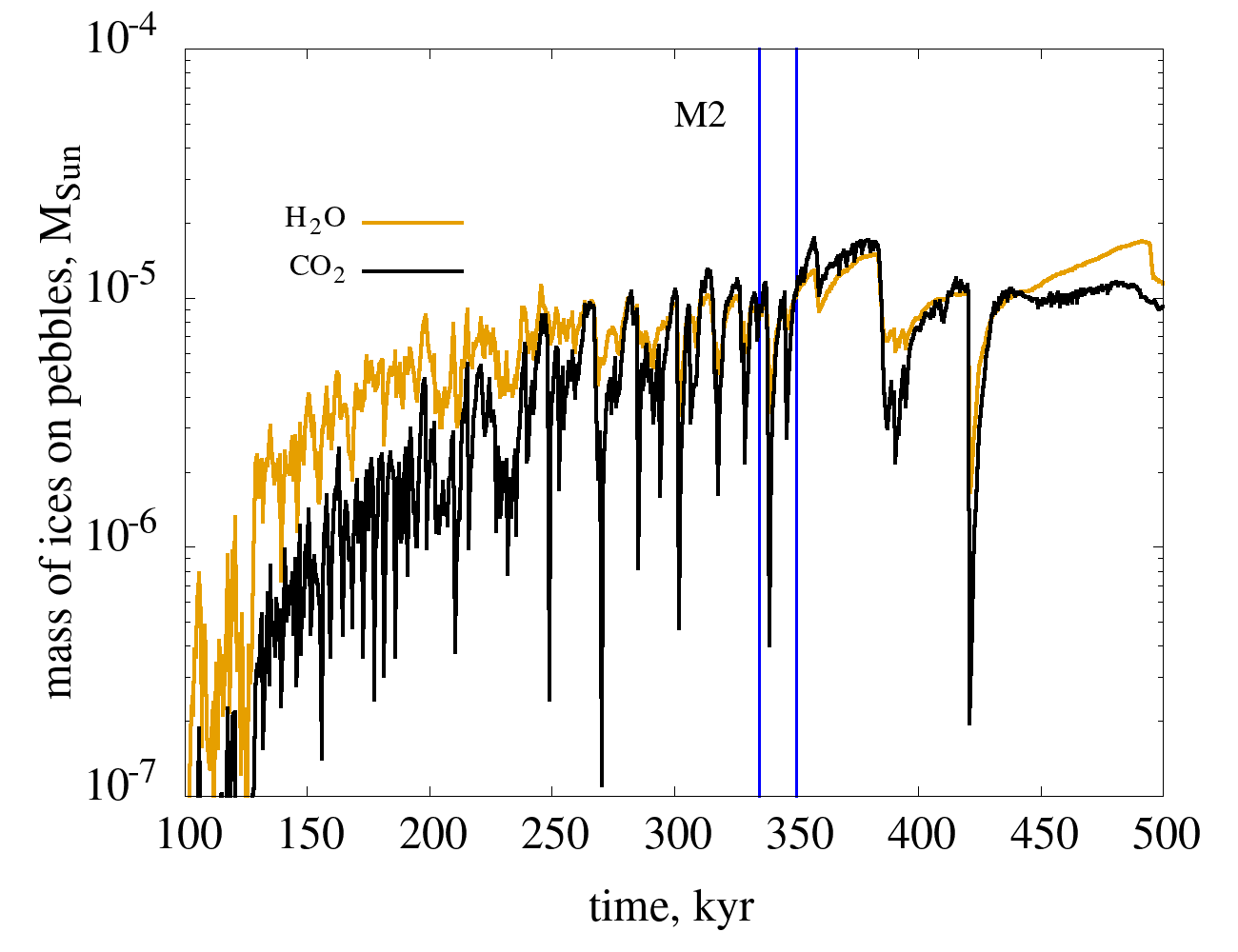}
\caption{Total disk-integrated mass of ices on pebbles versus time in models M1 (top), an outburst at 244.5 kyr and M2 (bottom), an outburst at $\sim340$ kyr. }
\label{ris:outburst_ices_on_pebbles}
\end{figure*}

Luminosity outbursts accompanying accretion rate increases heat the disk causing thermal ice desorption, shifting all volatile snowlines outward \citep{2016Natur.535..258C,Vorobyov2022,2023MNRAS.521.5826H}. This inevitably affects ice mantle composition on dust grains, including pebbles. Fig.~\ref{ris:outburst_ices_on_pebbles} shows temporal evolution of total mass of various ices on pebbles. At certain times ice masses drop sharply, corresponding to luminosity outbursts in Fig.~\ref{ris:accretion pattern}.

Some outbursts affect all ices, while others only reduce CH$_4$ and CO masses. These molecules are more volatile with lower desorption energies, their snowlines are farther from the star and do not always coincide with pebble regions \citep{2024MNRAS.530.2731T}. Their masses on pebbles are orders of magnitude lower than those of H$_2$O and CO$_2$. Thus even relatively weak luminosity outbursts can strongly affect their abundance in the disk if the snowlines lie in regions where stellar irradiation dominates heating rather than viscous heating.

\begin{figure*} 
\includegraphics[width =\columnwidth]{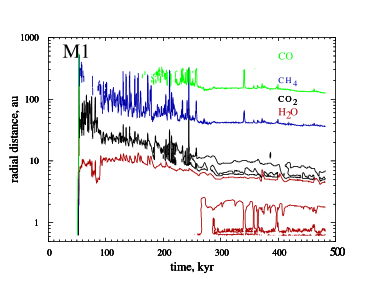}
\includegraphics[width =\columnwidth]{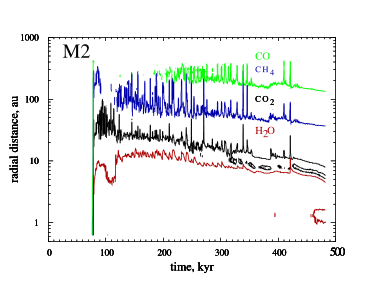}
\caption{H$_{2}$O, CO$_{2}$, CH$_{4}$ and $\rm CO$ snowlines for two models of different mass versus time.}
\label{ris:M!}
\end{figure*}

Fig.~\ref{ris:M!} shows the snowlines of the four molecules versus time. At certain times snowlines shift outward abruptly, corresponding to luminosity outbursts. For most weak outbursts the water snowline barely moves, because it lies in disk regions where turbulent viscous heating dominates over stellar irradiation. In contrast, CH$_{4}$, CO$_{2}$ and CO snowlines located at radii where stellar irradiation dominates heating show more pronounced shifts. This highlights the importance of analyzing radial boundaries where stellar irradiation heating dominates other heat sources for understanding the behavior of these molecules during outbursts \cite{Vorobyov2018A&A...614A..98V}.

At later times multiple water snowlines form in the disk. At $\approx1-2$\,au a region appears where water exists in ice phase. In model~M1 this occurs at $\approx250$\,kyr,
in model~M2 it happens after $\approx450$\,kyr. The formation of water ice in the inner region is associated with a $\approx1-2$\,au dust ring accumulating grown dust \citep[see][]{2024MNRAS.530.2731T}. This ring forms in the dead zone with the lowest $\alpha$-parameter and high gas/dust surface density. Consequently, heating is weakened there, causing relative temperature drops sufficient for water freeze-out.

Although the water snowline does not shift much and water ice mass on pebbles varies less than other molecules, water remains one of the most important molecules in protoplanetary disks. All disk pebbles are covered with water ice \citep{2024MNRAS.530.2731T}. This, combined with fragmentation velocity dependence on ice mantle presence (Section~\ref{sec:chem_model}), makes water the key pebble surface molecule deserving detailed consideration.

\begin{figure}
\includegraphics[width =\columnwidth]{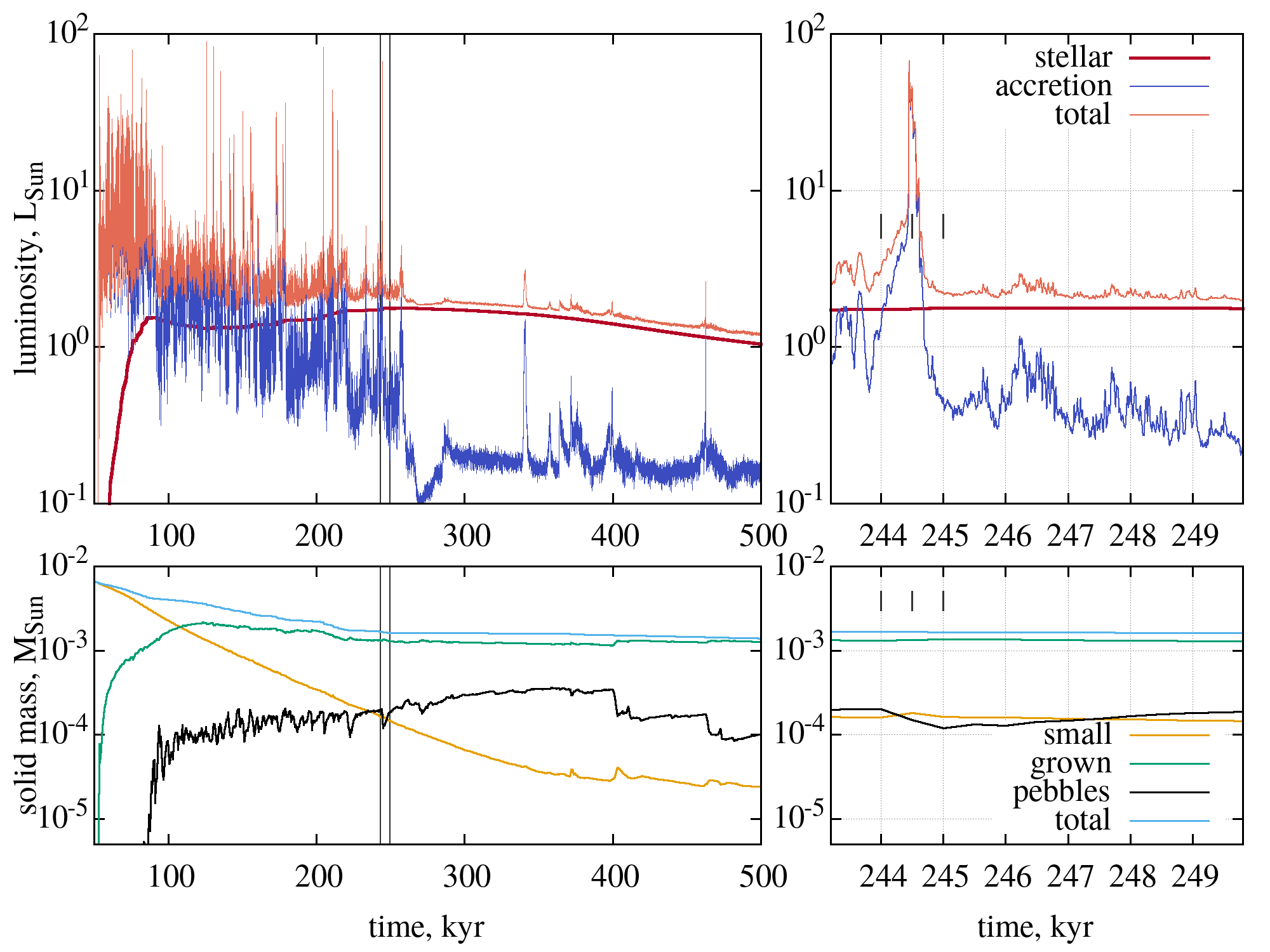}
\includegraphics[width =\columnwidth]{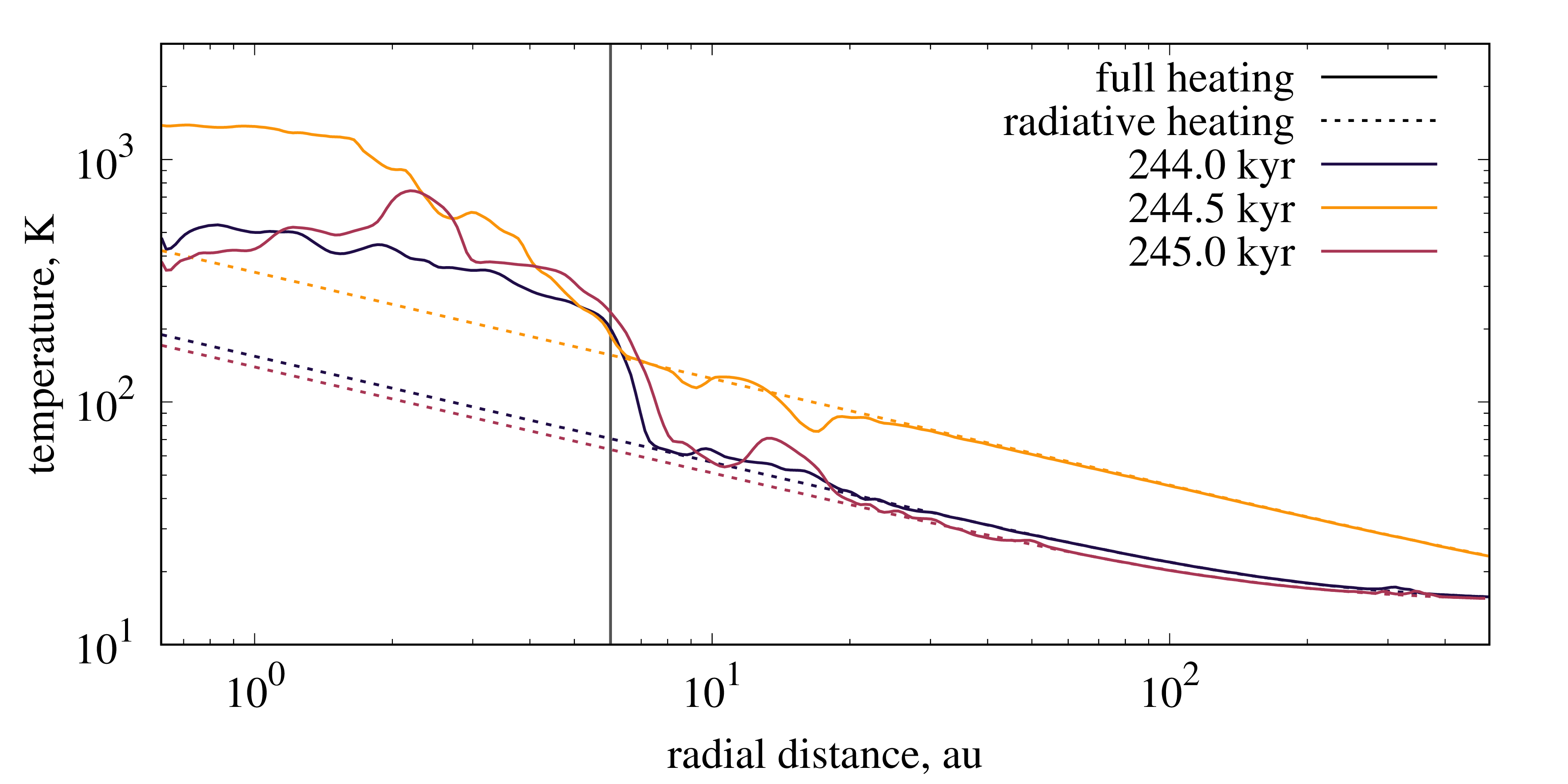}
\caption{Stellar luminosity and accretion luminosity (top panels) and total mass  of refractory components (middle panel) in model M1. Left panels show full disk lifetime, right panels show the 244.5~kyr outburst. Vertical lines mark the outburst on the left panels. Short vertical lines on the right panel correspond to times shown in Fig.~\ref{ris:outburst245_ices}. Bottom panel shows radial temperature profiles before, during and after the outburst. Solid lines show temperatures including all heating mechanisms, dashed lines show radiation-only temperatures. Black vertical line marks approximate water snowline position.}
\label{ris:outburst245}
\end{figure}

Let us examine outburst effects on pebbles and their composition in more detail using characteristic events in both models. In M1 this is the 244.5\,kyr event, in M2 the $\sim340$\,kyr event.

The first considered luminosity outburst profile and corresponding masses of refractory components are shown in right panels of Fig.~\ref{ris:outburst245}. Luminosity peaks at 244.5~kyr. The outburst lasts about two hundred years with $\approx70$\,$L_{\odot}$ magnitude. Pebble mass begins decreasing simultaneously with luminosity rise; by 245~kyr, i.e. after the outburst, it drops by roughly half, then recovers to pre-outburst values in $\sim4$~kyr. As pebble mass falls, small dust mass rises while grown dust mass changes only slightly. During the outburst pebbles are converted to small and grown dust. Grown dust also fragments to small dust in regions with no pebbles. Amount of grown dust increases from pebble destruction but it also fragments, so total dust mass changes weakly. In several thousand years around the outburst total dust mass shows slight decreases due to radial drift and accretion to the star.

This outburst is typical for early evolution in the M1 model. Similar changes in pebble mass occur during other outbursts, as seen in Fig.~\ref{ris:outburst245} lower left panel. During such outbursts mass of pebbles in the disk decreases by a factor of $\approx2$.

The luminosity outburst heats the disk, but temperature increases vary between disk regions depending on local heating mechanism contributions to thermal balance. This is illustrated in Fig.~\ref{ris:outburst245} bottom panel showing radial temperature profiles before, during and after the outburst. Pre- and post-outburst profiles are similar but both differ strongly from the outburst profile. Beyond $\approx7-8$~au midplane temperature $T_{\rm m.p.}$ approaches radiative temperature $T_{\rm irr}$, which is the temperature calculated assuming radiation is the sole heating source. In these regions external irradiation dominates heating, and luminosity outbursts have greatest impact, roughly doubling temperatures. Inside 7~au temperature exceeds radiative temperature by a factor of a few. Here  thermal structure is determined by viscous heating. Within $\approx 4$~au MRI develops, increasing $\alpha$, enhancing viscous heating and raising temperatures. Between $4-7$~au $\alpha$ does not change during the outburst and radiative heating contribution is small, so temperatures stays approximately the same. The outburst significantly heats all disk regions except $4-7$~au. This effect occurs for outbursts with amplitudes around a hundred solar luminosities.

\begin{figure*}
\includegraphics[width =2\columnwidth]{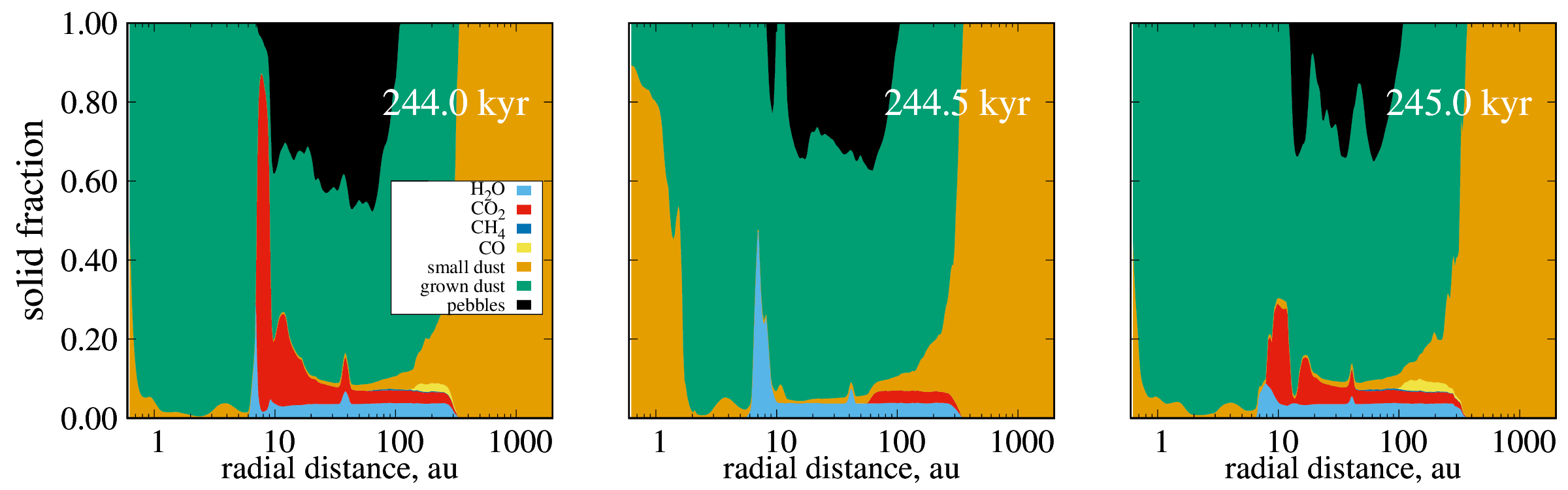}
\caption{Mass fractions of ices and refractory components relative to total mass of solid material in the disk before (left), during (center) and after (right) the luminosity outburst in model M1.}
\label{ris:outburst245_ices}
\end{figure*}

\begin{figure}
\includegraphics[width =\columnwidth]{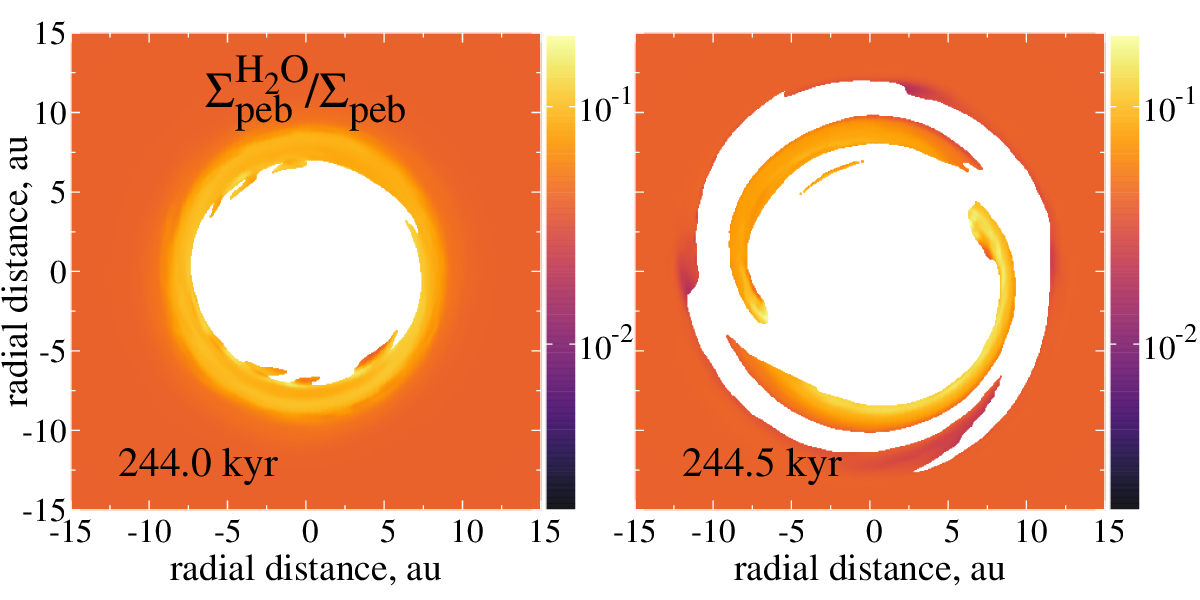}
\includegraphics[width =\columnwidth]{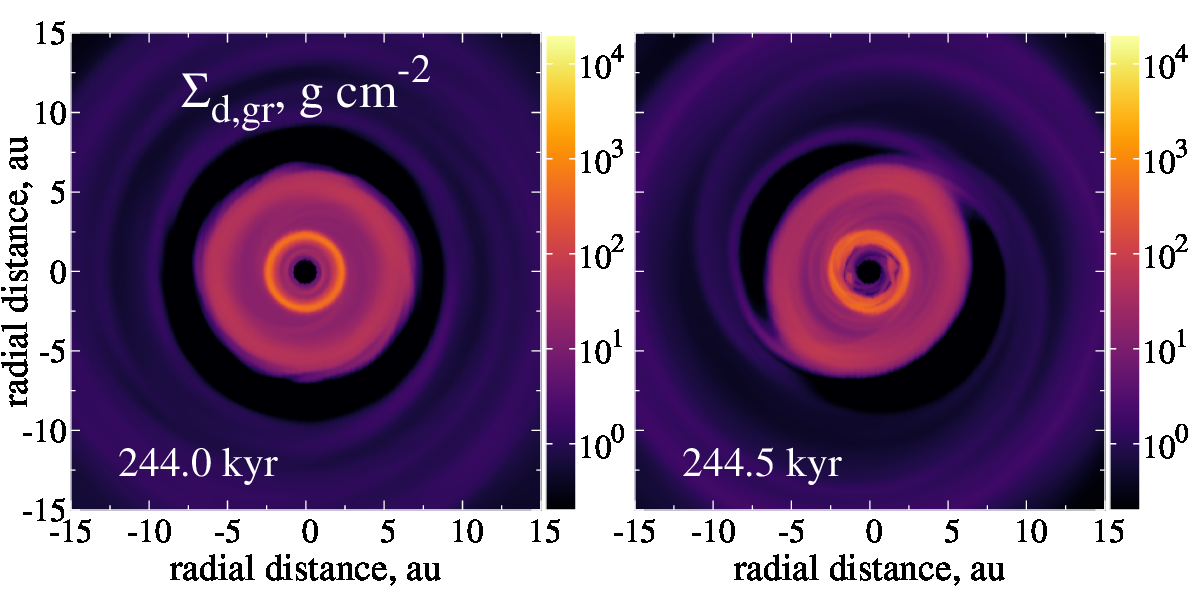}
\includegraphics[width =\columnwidth]{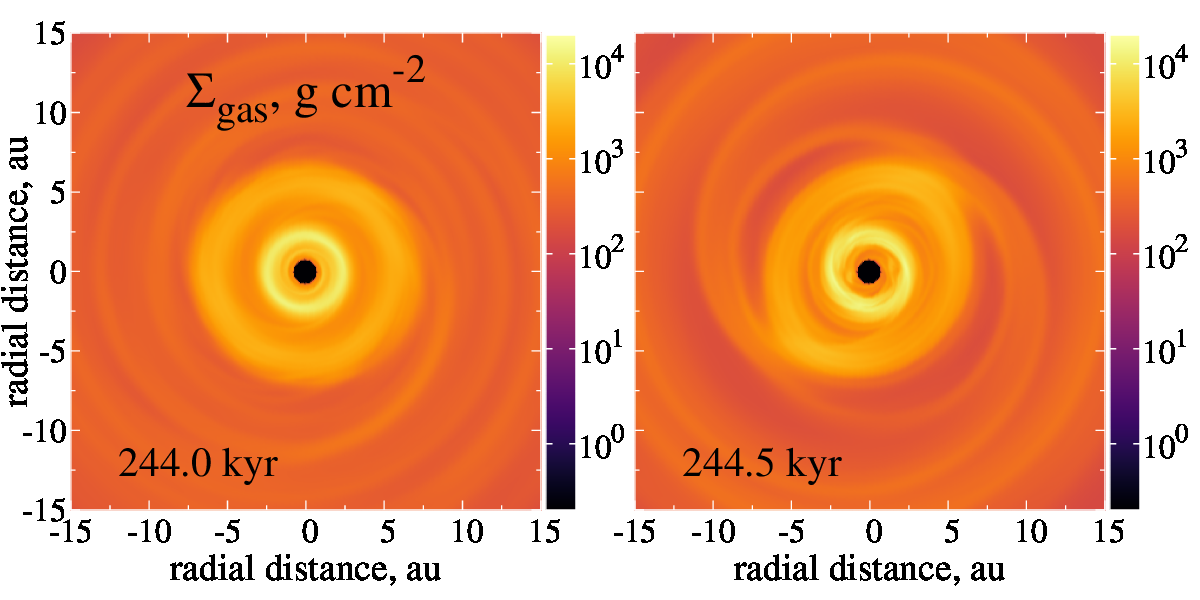}
\caption{Top panel: H$_2$O mass fraction on pebbles in the protoplanetary disk; middle panel: dust surface density; bottom panel: gas surface density, before and during the outburst in model M1.}
\label{ris:outburst_maps}
\end{figure}

Pebble mass reduction directly results from luminosity outbursts. Additional accretion luminosity heating causes thermal desorption of ice mantles across large disk areas. This is evident from significant snowline shifts in Fig.~\ref{ris:M!}. Near the water snowline on the inner side (closer to star), grains lose ice mantles to evaporation. Water desorption also occurs in spiral structures where temperatures are significantly elevated due to local gas heating. The zone where water does not desorb from dust includes not only regions outside the snowline but also spiral structures permeating the disk, which we examine in more detail below.

Grains in these regions are heated up and destroyed into monomers. This is also seen from corresponding increases in both grown and small dust masses from pebble fragmentation (we remind the reader that grown dust in our model is everything above one micron).
When outbursts end, volatiles return to grain surfaces and dust growth resumes. Post-outburst volatile freeze-out times can be estimated as inverse adsorption rates onto dust \citep[see equation (27) in][]{2024MNRAS.530.2731T}. Near water and CO$_2$ snowlines freeze-out times are at most days. In outer regions with CO and CH$_4$ snowlines, lower temperatures and dust surface densities can lead to freeze-out times up to decades.

Characteristic times for grains to return to pre-outburst states and become pebbles again are around thousands of years. They are determined by combined timescales of  volatile freeze-out and dust growth. Water freeze-out times in these relatively high-density regions are very short compared to outburst durations, not exceeding several years as shown by astrochemical modeling of outburst chemical impacts \citet{2019MNRAS.485.1843W}. \citet{2013A&A...557A..35V} showed that CO and CO$_2$ freeze-out times in surrounding envelopes are thousands of years due to lower densities and temperatures favoring slower freeze-out. Here post-outburst pebble formation times around a thousand years are set by dust growth rates. \citet{Vorobyov2022} showed that post-outburst dust growth times are shorter closer to the star, with typical values $\sim10^{3}$~years.

Fig.~\ref{ris:outburst245_ices} shows radial distributions of solid disk components before, during and after the considered outburst. The comparison between the left and the middle panel demonstrates thermal desorption effects from additional stellar accretion luminosity heating. During the 245~kyr outburst CO$_{2}$ fraction drops sharply over a wide radial range, while CH$_{4}$ and CO disappear completely. Their snowlines lie beyond $>7$\,au where accretion luminosity heating is important. Meanwhile the water snowline position does not change at all, remaining at 6\,au. It lies at radii where outburst heating is weak.

As Fig.~\ref{ris:outburst245_ices} shows, water ice radial distribution changes little with the outburst. Yet total disk water ice mass clearly decreases during outbursts, as seen in Fig.~\ref{ris:outburst_ices_on_pebbles}.
This apparent contradiction results from non-axisymmetric structure of the disk. Fig.~\ref{ris:outburst_maps} shows 2D~distributions of H$_2$O ice fraction on pebbles before and during the outburst. Water does sublimate between $\approx7-11$~au but not at all azimuths, leaving a spiral-shaped region where pebbles remain icy. The H$_2$O snowline lies at radii where viscosity and radiation compete for thermal structure dominance, leading to complex snowline geometry evolution. In regions where grains lose ice mantles, they immediately fall apart into monomers and their sizes decrease.
These regions contain about half of the total mass of pebbles in the disk. Pebbles are destroyed during the outburst as their ice mantles evaporate. This disruption is not visible in radially averaged distributions (Fig.~\ref{ris:outburst245_ices}) because regions with sublimated ice are averaged together with the regions having high H$_2$O ice fractions and pebble mass. Ice sublimation and pebble destruction during the outburst strongly depend on disk 2D~structure.

\begin{figure}
\includegraphics[width =\columnwidth]{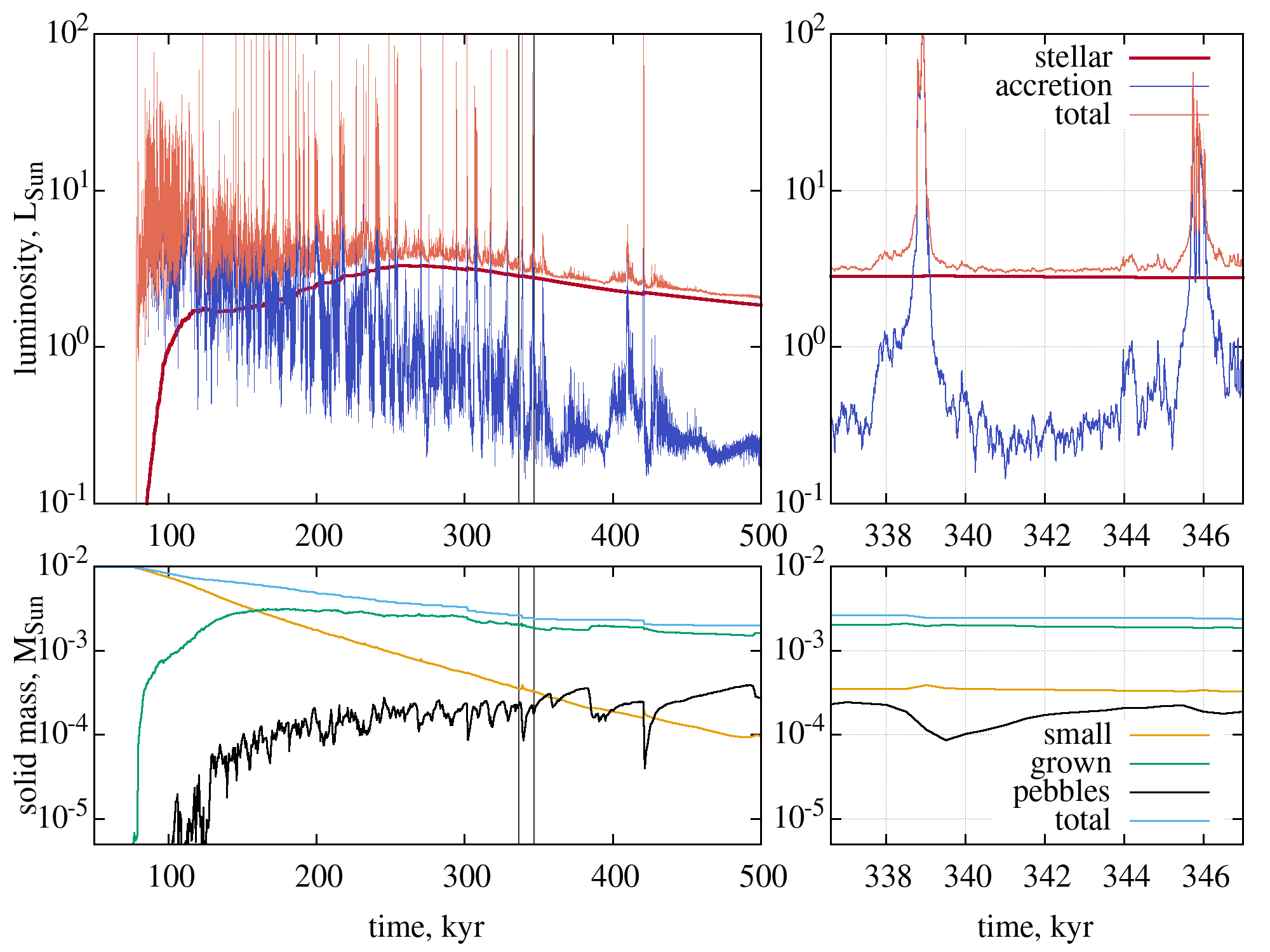}
\includegraphics[width =\columnwidth]{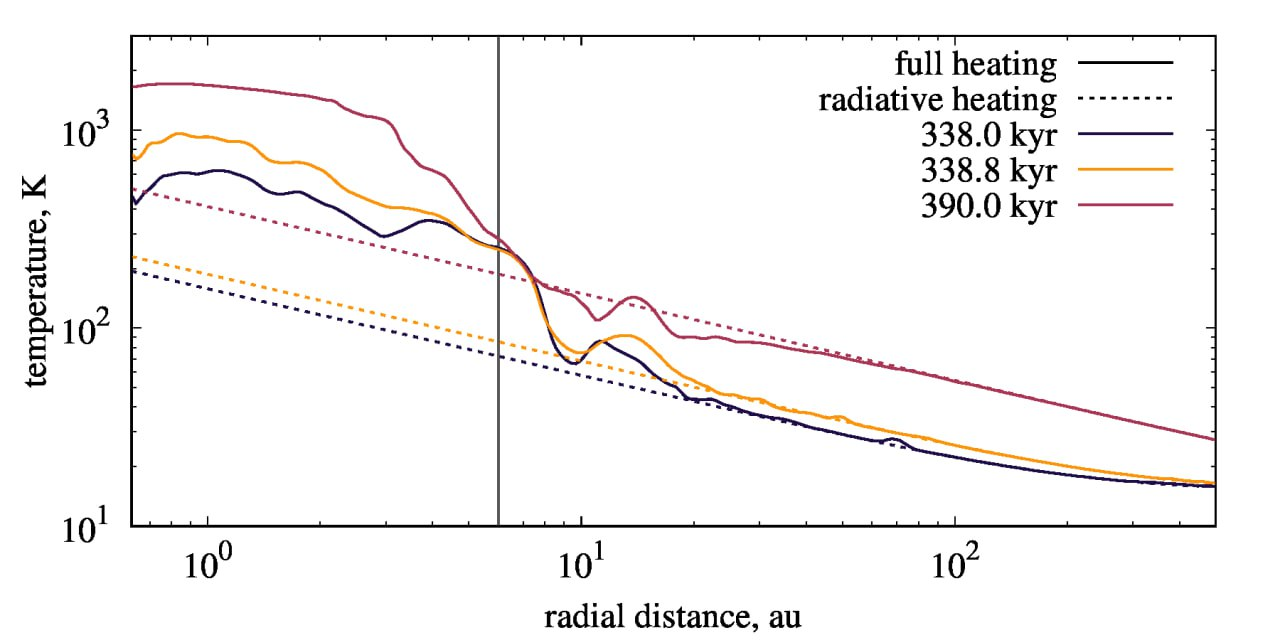}
\caption{Stellar luminosity, accretion luminosity and total mass of refractory components, plus radial temperature profiles before, during and after the 339 kyr luminosity outburst in model M2. Bottom panel shows radial temperature profiles before, during and after the outburst. Solid lines show temperatures including all heating mechanisms, dashed lines show radiation-only temperatures. Black vertical line marks approximate water snowline position.}
\label{ris:outbursts_massive_340}
\end{figure}

Luminosity outbursts like those above also occur in model M2. Like the M1 outburst, these are accompanied by pebble mass reductions caused by disruption of grains that lost their ice mantles into monomers and subsequent pebble recovery as water ice refreezes. Radial temperature profiles before, during and after the\,339 kyr outburst in Fig.~\ref{ris:outbursts_massive_340} show temperature changes similar to Fig.~\ref{ris:outburst245} bottom panel.

\section{Conclusions}
\label{sec:theend}
In this work we model formation and global evolution of a self-gravitating viscous protoplanetary disk using the 2D hydrodynamic FEOSAD code. We examine characteristics of self-consistent luminosity outbursts in the model and analyze their effects on pebbles and composition of their ice mantles in protoplanetary disks. Our main conclusions are as follows:

\begin{itemize}
\item During luminosity outbursts total pebble mass in the disk decreases by $\approx2\times$ due to desorption of ice mantles and destruction of dust grains into monomers.
Ice mantles favoring dust growth are essential for pebble existence in our model.
\item Pebbles recover within several thousand years after the outburst. Recovery timescales are set by dust growth during collisional coagulation, significantly exceeding post-outburst water freeze-out times (around several years), as previously shown in \cite{Vorobyov2022}.
\item Luminosity outbursts more strongly affect positions of CO$_2$, CH$_4$ and CO snowlines than the position of the water snowline. This occurs because the H$_2$O snowline lies in the viscous heating dominated region where temperatures are less affected by the luminosity outburst. Conversely, CO$_2$, CH$_4$ and CO snowlines reside in regions sensitive to stellar irradiation heating.
\item Ice mantle desorption proceeds substantially two-dimensionally due to the presence of spiral substructures during early disk evolution stages. The radial position of the water snowline shifts only slightly.
\end{itemize}

Special attention should be paid to differences between model results and FU~Ori object observations, particularly the FUor V883~Ori. According to \cite{Tobin2023} and \cite{Cieza2016}, V883~Ori outbursts cause significant radial shifts in water snowlines. This may indicate passive heating dominance in snowline regions of such objects, where temperature is mainly determined by central source radiation enhanced by outbursts.

Unlike V883~Ori, our young disk model shows little water snowline movement. This occurs because the H$_2$O snowline lies in the viscous heating dominated region where temperature is primarily set by local energy generation and transport processes, with stellar irradiation playing a secondary role. Such differences highlight the need for further study of active-to-passive heating transitions during protoplanetary disk evolution and the role of this transition in forming observable characteristics of an outbursting object.

\section*{Acknowledgments}
The authors thank the anonymous referee for constructive comments.
This research was supported by the Russian Science Foundation grant No. 22-72-10029, https://rscf.ru/project/22-72-10029 (Sections 1, 2.1, 2.2, 3, 4). Tamara Molyarova's work was supported by the Ministry of Science and Higher Education of the Russian Federation, State Assignment No. GZ0110/23-10-IF.

\bibliographystyle{mflds}
\bibliography{mflds}

\end{document}